\documentclass[sigconf,natbib=true,anonymous=false]{acmart}
\usepackage{amsmath,nccmath,tabularx} 
\usepackage{algorithm}
\usepackage{algorithmic}
\usepackage{color}
\usepackage{graphicx}
\graphicspath{{images/}}
\usepackage{wrapfig,lipsum}
\usepackage{tabularx}
\usepackage{graphicx}
\usepackage{lscape}
\usepackage{subcaption}
\usepackage{latexsym}
\usepackage{pifont}
\usepackage{multirow}

\usepackage{enumitem}
\setlist{leftmargin=3mm}
\usepackage[T1]{fontenc}
\usepackage[utf8]{inputenc}
\usepackage[font=small,labelfont=bf]{caption}

\DeclareMathOperator*{\argmax}{arg\,max}
\DeclareMathOperator*{\argmin}{arg\,min}
\DeclareMathOperator*{\cossim}{cos\_sim}

\AtBeginDocument{%
  \providecommand\BibTeX{{%
    \normalfont B\kern-0.5em{\scshape i\kern-0.25em b}\kern-0.8em\TeX}}}

\copyrightyear{2023} 
\acmYear{2023} 
\setcopyright{acmlicensed}\acmConference[SIGIR-AP '23]{Proceedings of the 1st International ACM SIGIR Conference on Information Retrieval in the Asia Pacific}{November 26--29, 2023}{Beijing, China}
\acmBooktitle{Proceedings of the 1st International ACM SIGIR Conference on Information Retrieval in the Asia Pacific (SIGIR-AP '23, November 26--29, 2023, Beijing, China}
\acmPrice{15.00}
\acmDOI{xxxxxx}
\acmISBN{978-1-4503-8037-9/21/07}

\begin{document}

\title{Selecting which Dense Retriever to use for Zero-Shot Search}




\author{Ekaterina Khramtsova}
\affiliation{%
	\institution{The University of Queensland}
	\streetaddress{4072 St Lucia}
	\city{Brisbane}
	\state{QLD}
	\country{Australia}}
\email{e.khramtsova@uq.edu.au}

\author{Shengyao Zhuang}
\affiliation{%
	\institution{The University of Queensland}
	\streetaddress{4072 St Lucia}
	\city{Brisbane}
	\state{QLD}
	\country{Australia}}
\email{s.zhuang@uq.edu.au}

\author{Mahsa Baktashmotlagh}
\affiliation{%
	\institution{The University of Queensland}
	\streetaddress{4072 St Lucia}
	\city{Brisbane}
	\state{QLD}
	\country{Australia}}
\email{m.baktashmotlagh@uq.edu.au}

\author{Xi Wang}
\affiliation{%
	\institution{Neusoft}
	\city{Shenyang}
	\country{China}}
\email{wxi@neusoft.com}

\author{Guido Zuccon}
\affiliation{%
	\institution{The University of Queensland}
	\streetaddress{4072 St Lucia}
	\city{Brisbane}
	\state{QLD}
	\country{Australia}}
\email{g.zuccon@uq.edu.au}

\begin{abstract}
We propose the new problem of choosing which dense retrieval model to use when searching on a new collection for which no labels are available, i.e. in a zero-shot setting.  Many dense retrieval models are readily available. Each model however is characterized by very differing search effectiveness -- not just on the test portion of the datasets in which the dense representations have been learned but, importantly, also across different datasets for which data was not used to learn the dense representations. This is because dense retrievers typically require training on a large amount of labeled data to achieve satisfactory search effectiveness in a specific dataset or domain. Moreover, effectiveness gains obtained by dense retrievers on datasets for which they are able to observe labels during training, do not necessarily generalise to datasets that have not been observed during training. 



This is however a hard problem: through empirical experimentation we show that methods inspired by recent work in unsupervised performance evaluation with the presence of domain shift in the area of computer vision and machine learning are not effective for choosing highly performing dense retrievers in our setup.
The availability of reliable methods for the selection of dense retrieval models in zero-shot settings that do not require the collection of labels for evaluation would allow to streamline the widespread adoption of dense retrieval. This is therefore an important new problem we believe the information retrieval community should consider.
Implementation of methods, along with raw result files and analysis scripts are made publicly available at \url{https://www.github.com/<anonymized>}

\end{abstract}

\ccsdesc[500]{Information systems~Evaluation of retrieval results}
\keywords{Model selection, Dense retrievers, Zero Shot Model Evaluation}

\maketitle

\section{Introduction}

Model analysis in zero-shot scenarios involves comparing and evaluating different models based on their ability to generalize to new datasets, that have not been seen during training. In many real-world settings, it may not be possible to obtain judgments for these new datasets (which we refer to as the target dataset), as labeling data can be expensive, time-consuming and sometimes even impossible (e.g., in medical or legal domain, access to labeled data can be restricted due to privacy concerns or regulations). Therefore, the reliance on standard evaluation practices and metrics  on these new dataset is not possible.

Furthermore, the data distribution in the target dataset is often different from the data distribution in the training dataset: this is also known as the domain shift problem. This can have a significant impact on the effectiveness of the model, as it may not be able to generalize to the new domain/target dataset~\cite{Chen2022OutofDomainST}. Therefore, judging the model based exclusively on its effectiveness on the training dataset might result in inaccurate predictions.

The issue of data distribution shifting has gained increasing attention in the field of information retrieval (IR). This is largely due to the recognition that dense retrievers (DRs) based on pre-trained language models (PLMs) have shown exceptional effectiveness on in-domain data. However, recent studies~\cite{thakur2021beir,ren2022thorough,zhuang2021dealing,zhuang2022char} have also highlighted the problem of domain shift for these DRs, whereby the effectiveness of a DR varies depending on the target domain, which may differ from the domain of the data it was trained on. This poses a significant challenge to the practical deployment of DRs in real-world applications.
Researchers in the IR community are focusing on developing novel pre-training or fine-tuning techniques to improve the zero-shot ability of DR models~\cite{gururangan-etal-2020-dont,wang2022gpl,ma2021zero,lin2023train}. However, the analysis of the leaderboard of a common zero-shot dense retrieval benchmark dataset, BEIR~\cite{thakur2021beir}, reveals that the effectiveness of different DR models varies across different domains/datasets, and no clear one-size-fits-all DR has emerged. Given the rapid development of new DR models, it is crucial to explore an alternative research direction. Rather than attempting to train a generalized DR model that performs well across all domains and datasets, we propose to develop methods that search engine practitioners can reliably use to select a DR model from an existing pool of DR checkpoints that performs best on their application's domain or dataset. This line of research is vital for the practical deployment of DRs in real-world scenarios.

In this paper, we introduce, formalize and operationalize the problem of model selection in  IR, which involves selecting the best DR model from a pool of trained models, or alternatively ranking the pool based on a criterion that is indicative of their effectiveness on the unlabeled target dataset. We demonstrate the soundness of the problem by illustrating that the effectiveness of a model varies across target datasets due to differences in data quality, the nature of domain shift, and the evaluation metric employed, as each metric prioritizes different aspects of model effectiveness. In other words, a model might perform well on some datasets, while failing on others. We further provide evidence that the performance of a model on the source dataset is not always a reliable indicator of its effectiveness on target datasets, highlighting the necessity for an alternative criterion specifically tailored to each target dataset. An example of such a criterion is the evaluation of the model's level of uncertainty, with the hypothesis that models with lower uncertainty are more likely to produce accurate predictions.


The goal of this paper is thus to provide an extensive overview of the current state-of-the-art methods for model selection, which have been mostly developed in the fields of computer vision and machine learning. We outline their limitations and summarize the main challenges in their adaptation to IR tasks. We further provide an outline of our findings along with associated reflections and challenges, and propose possible future directions for exploration. It is worth noting that while the problem of unsupervised model selection has been explored in the general machine learning field, there has been little to no exploration of this problem in the context of information retrieval.

\section{Related Work}

In this section, we will give an overview of the current state-of-the-art methods of unsupervised model selection and its relation to IR. 
In particular, we target the problem of selecting the best dense retriever model for the datasets from BEIR collection,
that contains a wide range of corpora from  diverse text retrieval tasks and domains for zero-shot evaluation.

\subsection{Unsupervised Model Selection}

The problem of Unsupervised Model Selection is actively researched in the context of general deep learning tasks.
It involves choosing the best model for an unsupervised target dataset, that belongs to a distribution different from the original training dataset.    

Corneanu et al.~\cite{Corneanu2019} analyze various topological statistics, calculated from the correlation matrix of the network's activations, and show that they are correlated with model generalizability.

You et al.~\cite{pmlr-v97-you19a} introduce Deep embedded validation (DEV)- a method to obtain estimations of the target risk by assessing the likelihood of each validation sample to belong to the target domain. 
The authors show that the small value of DEV correspond to a large generalizability of the model and can be therefore used for selecting the best performing model for the dataset in hand.

Soft Neighborhood Density method (SND) was proposed by Saito et al.~\cite{9711410} to evaluate the generalisability of the classifier networks by measuring the quality of  clustering via a relative distance of the neighboring samples. The intuition is that the desired model should encode target samples of the same class into dense neighborhoods. 
Another method of cluster evaluation, called Topological Uncertainty,  was proposed by  Lacombe et al.~\cite{Lacombe2021TopologicalUM}. For each class (or label), the authors create a prototype of the topological activation footprint, and evaluate how different is the embedding of each target sample from its closest prototype.
Our retrieval task is drastically different from classification: in our case, well-performing  model does not necessarily generate well-separated clusters in their embedding space. 


Neural Persistence measure, proposed by Rieck et al.~\cite{Rieck19a}, compares the networks based on the topological features of their last fully connected layers. Despite showing a correlation with model generalizability, Neural persistance-based ranking will provide the same result for all the datasets, as it is purely weight-based and dataset-independent. Additionally, for a fair comparison, Neural Persistence requires the same layer dimensionality across models, which is not the case in our experimental setup.

\subsection{Unsupervised OOD Performance Evaluation}

The problem of unsupervised Out-Of-Ditribution (OOD) performance evaluation involves estimating the performance of a model with the presence of domain shift between the train and test data. Note that the experimental setup of performance estimation is slightly different to model selection: here, only one model is given, and the task is to predict its performance on several target datasets.

The first group of methods analyzes the quality of model predictions through a range of metrics, calculated from the network output. For example, Miller et al.~\cite{Miller2021AccuracyOT} show a positive correlation between in-domain and out-of-domain performances (Method 1 in our analysis); 
Hendrycks et al.~\cite{hendrycks17baseline} use the statistics of softmax outputs to identify misclassifications; 
Guillory et al.~\cite{Guillory2021PredictingWC} evaluate the performance of the models based on the uncertainty of its predictions, approximated via the difference of confidences between the base dataset predictions and the target dataset predictions.  
Garg et al.~\cite{ATC}  estimate the average confidence threshold (ATC) from the validation data, above which the prediction of the network is considered to be incorrect.

The second group of methods evaluate the quality of the model embeddings. For example, Deng et al.~\cite{Deng2021WhatDR} show negative correlation between recognition accuracy and the Frechet distance between network activations to the source and the target datasets; while Jiang et al.~\cite{jiang2018predicting} predict generalization gap using margin distribution - the distances of training points to the decision boundary. Margin distribution requires class predictions, while Frechet distance only relies on hidden embeddings, that is why we use Frechet distance in our analysis (see Method 3). 

The last group of methods for performance evaluation focuses on behaviour analysis of the network. For example, Bridal et.al. ~\cite{Birdal2021IntrinsicDP} analyze training dynamics of the model; in particular, the authors measure generalization of a model via the topological properties of its optimization trajectories.
Instead of monitoring the network during training, the authors of \cite{chuang2020estimating, Jiang2022AssessingGO, Chen2021DetectingEA} perform the training of the same network several times, and measure the disagreement between the resulting trained models on the target dataset.     
Another work by Deng et al.~\cite{Deng2021WhatDR} shows how the performance on auxiliary task can be used to estimate the performance on the main task. 
Finally, \citet{khramtsova} propose to analyze how the network changes when it is fine-tuned on the target dataset with an unsupervised loss (e.g., entropy minimization). The authors show that the degree of change in network weights is negatively correlated with the effectiveness of the network on target dataset.



\subsection{Zero-shot Dense Retrieval}

Bi-encoder dense retrievers, initialized with pre-trained language models (PLMs) and fine-tuned with supervised data, have shown remarkable effectiveness in in-domain information retrieval tasks~\cite{zhao2022dense,yates2021pretrained}. However, recent studies on the BEIR benchmark dataset~\cite{thakur2021beir,ren2022thorough} have revealed that DRs suffer from the domain shift problem: the effectiveness of a DR varies depending on the corpus domain, which can be different from the domain of the data it was trained on.

The most straightforward domain adaptation method for PLM-based DRs is to first continue pre-training the PLMs on the target domain corpus before fine-tuning them with labeled data~\cite{gururangan-etal-2020-dont}. However, it is often difficult, and costly, to obtain sufficient in-domain labeled training data for IR tasks. Furthermore, it is often the case that DRs must be deployed in a zero-shot setting on a target corpus, which presents additional challenges for domain adaptation. One simple solution for this issue is to pre-train on the target domain data using unsupervised training tasks, followed by fine-tuning on the source domain data where there is sufficient labeled training data~\cite{wang2022gpl}.

Another line of work aimed at improving the zero-shot ability of dense retrievers (DRs) is through \textit{query generation}. This technique has been widely used for enhancing retrieval effectiveness on in-domain data~\cite{nogueira2019doc2query}, and recent studies have shown its effectiveness in the BEIR zero-shot dataset as well~\cite{thakur2021beir,bonifacio2022inpars,jeronymo2023inpars}. In one approach, a query generator trained on general domain data is employed to synthesize domain-targeted queries for a target corpus, on which a dense retriever is trained from scratch~\cite{ma2021zero}. This method has also yielded promising results and has been utilized as a post-training method for adapting powerful MS MARCO retrievers to target domains~\cite{thakur2021beir}. More recently, \citet{wang2022gpl} proposed GPL, which further improves the DR's domain shift ability by combining a query generator with pseudo labeling from a cross-encoder.

However, our paper's aim is fundamentally different from the works mentioned in this section, where the focus is on enhancing the effectiveness of DRs on domain transfer tasks. Instead, we focus \textit{on selecting the best-performing DR model} on the target corpus from a pool of existing models.

\subsection{Query Performance Prediction}
In this paper we are interested in predicting the relative effectiveness of a DR model among a pool of available DR models. This problem shares several common aspects with the well known information retrieval task of query performance prediction (QPP)~\cite{he2006query,zhou2007query,raiber2014query,hauff2008survey,carmel2012query}. The goal of query performance prediction is to determine which queries, among a set pool of queries, a target system will perform best/worst on. The task we examine instead, is to determine which dense retriever model performs best/worst among a pool of dense retrievers. Nevertheless, evaluation practices in QPP can be adapted to our task, for example the measurement of the Kendall Tau correlation between two query rankings, popular in QPP works, can be adapted to our settings -- we in fact do this in Section~\ref{sec:eval_measures}. We further note that other measures developed to evaluate QPPs, e.g., the $\tau_{AP}$ measure~\cite{yilmaz2008a-new-rank}, can also be adapted to our task, but we leave this to future work.
\section{Model Selection for Dense Retrievers}

In this section we first introduce the formalization of the problem of model selection for dense retrievers. We then discuss the challenges of model selection task, specific to DRs. We finally provide a description of the methods we investigate for model selection and how they are adapted within the context of DRs.

\subsection{Problem Formulation}

Consider a set of rankers $\mathcal{R} = \{R_1, R_2, \ldots R_n\}$ and the source dataset $\mathcal{S}$, which was used for training the rankers. Each ranker represents a dense retriever (dual- or bi- encoder model) $R=\{E_Q, E_D\}$, that separately encodes the query and the document into dense vectors. The relevance score between the query and the document is computed using a similarity function. In this work, we consider dense retrievers that use either cosine similarity or dot product:  
\begin{equation}
	dot(q, d) = E_Q(q)^T E_P(p); \hspace{8pt} \cossim(q, d) = \dfrac{E_Q(q)^T E_D(d)}{||E_Q(q)|| ||E_D(d)||} \nonumber
\end{equation}

Consider further a target dataset $\mathcal{T}$ for which no labeled data was observed during the training of the rankers in $\mathcal{R}$. The target dataset $\mathcal{T}$ consists of a set of queries $Q_\mathcal{T}$ and a set of documents $D_\mathcal{T}$, i.e. $\mathcal{T} = \{Q_\mathcal{T}, D_\mathcal{T}\}$. In addition, a set of relevance judgments (also called labels) is denoted as $\mathcal{J}_{Q_\mathcal{T},D_\mathcal{T}}$; below we use $\mathcal{J}$ as a shorthand for $\mathcal{J}_{Q_\mathcal{T},D_\mathcal{T}}$ when this cannot be confused. The elements of this set are pairs that establish a relationship between the elements in $Q_\mathcal{T}$ and those in $D_\mathcal{T}$. Practitioners can construct an ordering of the rankers in $\mathcal{R}$, or a ranking of models, according to a target evaluation measure $\mathcal{E}$ by applying each ranker to the dataset and use the relevance judgments to compute the evaluation measure. Then, the rankers would be ordered in decreasing value of $\mathcal{E}$, forming the true ranking of rankers $\mathcal{O(R, T, E,J)}$ .

\begin{definition}[Zero-shot Model Selection]
The problem of model selection consists of predicting the ranking $\mathcal{O(R, T, E, J)}$ \textit{without accessing the relevance judgments} $\mathcal{J}_{Q_\mathcal{T},D_\mathcal{T}}$. This is equivalent to producing a ranking $\mathcal{\hat{O}(R, T, E)}$ of the rankers in $\mathcal{R}$ for dataset $\mathcal{T}$ and evaluation measure $\mathcal{E}$, such that $\mathcal{\hat{O}(R, T, E)}$ corresponds to the true ranking $\mathcal{O(R, T, E, J)}$. Note that $\mathcal{\hat{O}(R, T, E)}$ does not include the relevance assessments $\mathcal{J}_{Q_\mathcal{T},D_\mathcal{T}}$ as input.
\end{definition}

Given the problem of zero-shot model selection, we are interested in devising a method $\mathcal{M(R, T, E)}$ that produces the ranking $\mathcal{\hat{O}(R, T, E)}$. 

\begin{definition}[Methods for Zero-shot Model Selection]
A method $\mathcal{M(R, T, E)}$ for zero-shot model selection takes as input the set of rankers $\mathcal{R}$, the target dataset $\mathcal{T}$ and the evaluation measure $\mathcal{E}$ and produces the predicted ranking $\mathcal{\hat{O}(R, T, E)}$.
\end{definition}

The effectiveness of the model selection method can be measured by its correlation with the ground truth performance across various target datasets. A high correlation, denoted as $corr(\mathcal{\hat{O}(R, T, E)},$ $ \mathcal{O(R, T, E, J)})$, indicates that the method can closely approximate the ground truth ranking without relying on target labels $\mathcal{J}$.

Note that the problem of zero-shot model selection can be relaxed to allow for the prediction methods to access a subset of the relevance judgments $\mathcal{J}_{Q_\mathcal{T},D_\mathcal{T}}$. For example, this problem could be revisited in the context of \textit{few-shot retrieval}, where the model selection method (and possibly but not necessarily also the rankers in $\mathcal{R}$) have access to a subset $F \in \mathcal{J}_{Q_\mathcal{T},D_\mathcal{T}}$ of all relevance judgments available for the dataset. We do not consider this setting in this paper.

\subsection{Challenges of model selection in IR} \label{subsection:challenges}

Before proceeding with the formalization of the chosen model selection methods, it is important to emphasize the challenges  associated with adapting existing approaches to the information retrieval setting. We demonstrate that the specificity of the field prevents a straightforward adaptation of many existing methods.  This reinforces our reasoning for choosing the final set of methods for comparison.

\begin{enumerate}[series=edu]
    \item \textbf{Variance in network structure: dense retrievers differ in types and numbers of network layers.}
\end{enumerate}

The machine learning research community lacks a consensus on the way a set of models for model selection is constructed. As a result, researchers employ different techniques to generate a pool of models, such as varying training hyper-parameters \cite{9711410}, taking different subsets of data for training \cite{Lacombe2021TopologicalUM}, changing initialization strategies \cite{Rieck19a} or using the combination of these techniques \cite{Miller2021AccuracyOT}. However, in all of these scenarios, the models' structure and depth remains the same, which significantly simplifies the task of model selection.

However, in this paper, we focus on a more realistic scenario for model selection in IR. In particular, instead of artificially synthesizing a collection of models and deriving a measure for early stopping \cite{Rieck19a}  or for hyper-parameter search \cite{9711410}, we aim to select the best model for the target dataset among the latest available state-of-the-art dense retrievers. As an example, consider the scenario of a hospital, that has a small highly specialized dataset in hand. With the appearance of a new model, that outperforms the other known baselines on the source dataset (e.g., MS MARCO), the question arises: Will the new model perform similarly well on the hospital's local unlabeled dataset? 

In practice, we use models from the leaderboard for zero-shot learning from the BEIR benchmark \cite{thakur2021beir}, that includes models with different architectures (Bert-based; DistilBert-based, Roberta-Based). Therefore, in order to meet the realistic demands for model selection, an essential additional constraint is that the model selection method must be insensitive to the architecture of the models.

\begin{enumerate}[resume*=edu]
    \item  \textbf{Variance in scoring function: based on the training procedure, document relevance is estimated by different scoring functions, e.g., cosine similarity or dot product.}
\end{enumerate}

While classifier models produce a unified output that represents the probability of the input sample belonging to a certain class, dense retriever scores do not represent probabilities. This difference prevents the direct adaptation of several methods \cite{pmlr-v97-you19a,hendrycks17baseline,Guillory2021PredictingWC,jiang2018predicting} to dense retrieval model selection.

Comparing models directly by the score is also not possible due to the variation in score range based on the score function used (for instance, cosine similarity with scores ranging from -1 to 1 vs. unbound dot product). Simultaneously, substituting one scoring function with the other is inadvisable as it results in a different ranking result. To mitigate this, we propose to use cosine similarity solely for model selection, while preserving the original score function for retrieval (see Method 2 in the following section for more details).

The final score-related challenge is that not only do score distributions differ among various models, but they also vary across different queries within a single model. It implies that unlike ATC~\cite{ATC},  there is no unified threshold across queries, below which the document is more likely to be irrelevant.

\begin{enumerate}[resume*=edu]
    \item \textbf{Large number of both network parameters and source dataset samples, which makes re-training impractical.}
\end{enumerate}

Several existing approaches require access to the training process, either for generating training trajectories from several consecutive training checkpoints \cite{Birdal2021IntrinsicDP}; or for evaluating the disagreement in judgments between similar models \cite{chuang2020estimating, Jiang2022AssessingGO, Chen2021DetectingEA}. However, retraining dense retrievers require significant computational resources and expertise. In addition, if the code was not made publicly available, it can be difficult to reproduce the exact training conditions, including hyper-parameters and data pre-processing steps. Therefore, in this work, we focus on methods that do not require re-training on the source dataset.

    

    

\subsection{Method 1: In-Domain Performance}\label{method_1}

A most naive approach is to rank models by their in-domain performance, or the performance on the source dataset. Let $e_S(R)$ be an evaluation measure of a ranker $R$ on the source dataset $S$. Then, the In-Domain Performance-based model ranking is defined as follows: 
\begin{equation}
	Method_1 = argmax_{R\in\mathcal{R}}(e_S(R))
\end{equation}

\subsection{Method 2: Query Similarity}\label{method_2}
Query Similarity Score evaluates a ranker by the proximity of its source and target query representations. The premise behind this score is that the most generalizable ranker should produce similar embeddings for both source and target queries.

Let target query relevance score - $tqr$ -  be the cosine similarity between a target query and its closest counterpart within source queries:
\[tqr(q_t, Q_S) = \argmax_{q_s \in Q_S} (\cossim(q_t, q_s))\]

Then the Query Similarity Score of a ranker $R$ is the average target query relevance score across target queries:

\begin{equation}
	Q\_sim(Q_S, Q_T) = \dfrac{\sum_{q_t \in Q_T}{tqr(q_t, Q_S)}}{|Q_T|} 
\end{equation}

Finally, the Query Similarity Score-based model ranker is defined as follows: 
\begin{equation}
	Method_2 = argmax_{R\in\mathcal{R}}(Q\_sim_R)
\end{equation}

\subsection{Method 3: Corpus Similarity}\label{method_3}
Corpus Similarity Score evaluates the similarity of corpus representations. The intuition is that the ranker should generalize well if the source and the target corpora are encoded in a similar way.

Using per-sample cosine similarity from the previous section proves to be impractical for corpus comparison due to a high dimensionality of their representations. Instead, we propose to employ Fréchet distance~\cite{DOWSON1982450} between network activations of the source and target corpora. Deng and Zheng \cite{deng2020labels} show that small Frechet distance between network activations is correlated with the high recognition accuracy of convolutional classifiers.   

Let $\mu_s, \mu_t, \Sigma_S, \Sigma_T$ be the mean and the covariance matrix of the network representations, produced by the source and the target corpora:

\[\mu_s = \dfrac{\sum_{d_s \in D_S}{E_D(d_s)} }{|D_s|}; \mu_t = \dfrac{\sum_{d_t \in D_t}{E_D(d_t)} }{|D_t|} \]

\[\Sigma_S = \dfrac{ \sum_{d_s \in D_S}{[E_D(d_s)- \mu_s][E_D(d_s)- \mu_s]^{T}} }
                   { |D_S|-1}\]

\[\Sigma_T = \dfrac{ \sum_{d_t \in D_T}{[E_D(d_t)- \mu_t][E_D(d_t)- \mu_t]^{T}} }
                   { |D_T|-1}\]

Then, Fréchet distance between source and target corpus representations is defined as following: 

\begin{equation}
	FD(D_S, D_T) = (||\mu_s - \mu_t||)^2 + Tr(\Sigma_S + \Sigma_T - 2 Tr \sqrt{\Sigma_S \Sigma_T})
\end{equation}

Finally, the Corpus Similarity-based model ranker is defined as follows: 

\begin{equation}
	Method_3 = argmin_{R\in\mathcal{R}}(FD_R)
\end{equation}

\begin{table*}[t!]
    \resizebox{\textwidth}{!}{
    \begin{tabular}{l|cc|cc|ccccccccccccccc}
        \toprule
    Model              & \multicolumn{2}{c|}{In-dom.}& \multicolumn{2}{c|}{Zero-shot} & \multicolumn{10}{c}{Zero-Shot Datasets}   \\
                    & Rank  &  MS MARCO     & Rank         & Avg.        & COVID         & NF & NQ    & Hotpot & FiQA  & ArguAna & DBPEDIA & SCIDOCS  & SciFact &  Quora\\
    \midrule
    Contriever         & 4 & 0.407                & 1            &\textbf{ 0.496}               & 0.596                & \textbf{0.329}    & 0.498 & \textbf{0.638}    & \textbf{0.329} & \textbf{0.446}   & \textbf{0.413}   & \textbf{0.165}   & \textbf{0.677}   & \textbf{0.865}    \\
    CoCondenser        & 2 & 0.433                & 2            & 0.472               & \textbf{0.752 }               & 0.297    & 0.495 & 0.562    & 0.297 & 0.377   & 0.364   & 0.137   & 0.556   & 0.856    \\
    DistilBERT (TAS-B) & 3 & 0.408                & 3            & 0.459               & 0.481                & 0.319    & 0.463 & 0.584    & 0.300 & 0.427   & 0.384   & 0.149   & 0.643   & 0.835    \\
    SimLM			   & 1 & \textbf{0.458}                & 4            & 0.443               & 0.527                & 0.318    & \textbf{0.502} & 0.568    & 0.297 & 0.376   &0.351    & 0.137   & 0.559   & 0.796    \\
    ANCE               & 7 &0.388                & 5            & 0.426          & 0.653                & 0.236    & 0.444 & 0.451    & 0.295 & 0.419   & 0.281   & 0.122   & 0.511   & 0.852    \\
    DistilBERT v3      & 5 & 0.389                & 6            & 0.425         & 0.477                & 0.256    & 0.450 & 0.513    & 0.257 & 0.426   & 0.338   & 0.133   & 0.538   & 0.855    \\
    DistilBERT (dot)   & 6 & 0.389                & 7            & 0.418          & 0.633                & 0.269    & 0.442 & 0.477    & 0.253 & 0.329   & 0.315   & 0.114   & 0.515   & 0.833    \\
    MiniLM-L-12        & 8 & 0.385                & 8            & 0.403          & 0.473                & 0.252    & 0.422 & 0.456    & 0.240  & 0.407   & 0.307   & 0.113   & 0.503   & 0.854    \\
    BERT-DPR           & 10 & 0.364                & 9            & 0.398               & 0.619                & 0.216    & 0.442 & 0.454    & 0.216 & 0.354   & 0.297   & 0.111   & 0.452   & 0.792    \\
    MiniLM-L-6         & 9 & 0.379                & 10           & 0.395          & 0.479                & 0.255    & 0.394 & 0.448    & 0.231 & 0.394   & 0.292   & 0.116   & 0.495   & 0.845    \\
    DistilBERT v2      & 11 & 0.336                & 11           & 0.364               & 0.242                & 0.258    & 0.362  & 0.427   & 0.212 & 0.429   & 0.288   & 0.133   & 0.495   & 0.815    \\
    
    \bottomrule   
    \end{tabular}
    }
    \caption{Effectiveness of the considered Dense Retrievers on the in-domain dataset (MS MARCO, from which the training data was also drawn) and on the BEIR zero-shot target datasets. \label{tbl:DR-eval}}
    \vspace{-2ex}    
\end{table*}

\subsection{Method 4: Extracted Document Similarity}\label{method_4}
A particularity of IR setup in comparison to other ML fields is a large dataset size: target corpora, and in particular source corpora often consist of millions of documents. For that reason, comparing full corpus representations might lead to over-generalization and, as a consequence, result in an inaccurate metric. 
Instead, we propose to adapt Method~\ref{method_3} to IR task by only comparing the subset of documents, extracted by the target query from the source corpus and from the target corpus. 

Let $l_S^q = [d_1, d_2, \ldots, d_k], d_i \in S$ be a list of $k$ documents, extracted by a ranker with a query $q$ from a source dataset $S$. In addition, let $l_T^q= [d_1', d_2', \ldots, d_k'], d_i' \in T$ be a list of $k$ documents, extracted with the same query from a target dataset $T$. 
Then, for a target query $q \in Q_T$, Fréchet distance between extracted subsets of source and target document representations is denoted as $FD(l_S^q, l_T^q)$. 

We average the resulting distances across target queries to obtain an adapted version of Fréchet distance:

\begin{equation}
	FD\_IR(D_S, D_T)= \dfrac{\sum_{q \in Q^T } FD(l_S^q, l_T^q) }{|Q^T|} 
\end{equation}

Finally, the Extracted Document Similarity-based model ranker is defined as follows: 
\begin{equation}
	Method_4 = argmin_{R\in\mathcal{R}}(FD\_IR_R)
\end{equation}

\subsection{Method 5: Binary Entropy}\label{method_5}
In the context of classification tasks, entropy is commonly employed to measure the level of uncertainty in a model's predictions. However, in our scenario, it cannot be directly applied due to the possibility of multiple relevant documents per query. To adapt the entropy-based method for our task, we apply a two-step process. First, we transform the scores, produced by the dense retriever, into probabilities indicating the relevance of each document to the query. Next, we compute binary entropy for each document in the rank. We then calculate the probability-at-rank distribution by adding up the binary probabilities for all the documents in the ranking.


Let $s_T^q = [s_1, s_2, .., s_k]$ be a list of scores, obtained by a ranker with the query $q$ from the target dataset $T$; scores can be either a dot product or a cosine similarity, depending on the model. We first need to normalize the scores. For that,  we mine for the negative samples as follows: $\tilde{s}_T^q = [\tilde{s}_1, \tilde{s}_2, ..., \tilde{s}_{100}]$; $\tilde{s}_T^q \cap s_T^q = \emptyset$. Then we approximate the minimum score for the rank and normalize the scores to get probabilities: $min_T^q = \argmin (\tilde{s}_T^q)$; $p(d_i) = (s_i-min_T^q) /(s_1 - s_i)$

Assume that $p(d_i)$ is the probability that the document at rank $i$ is relevant, while $p(d_1, \ldots, d_k)$ is a probability distribution over the relevance associated with a document list of length $k$ (i.e., probability-
at-rank). Then, the binary entropy  is:
\begin{equation}
	H(p(d_i)) = - p(d_i) \log p(d_i) - (1- p(d_i)) \log (1-p(d_i))
\end{equation}

The entropy probability-at-rank distribution for query $q$ is therefore defined as follows:

\begin{eqnarray}
	H^q(p(d_1, \ldots, d_k)) = \sum_{i=1}^{k} H(p_i)\\	
	H_R = \sum_{q \in Q^T }{H^q} \mathbin{/}  |Q^T|
\end{eqnarray}

Finally, the Binary Entropy-based model ranker is defined as follows: 
\begin{equation}
	Method_5 = argmin_{R\in\mathcal{R}}(H_R)
\end{equation}

\subsection{Method 6: Query Alteration}\label{method_6}

This method aims to rank models by assessing their ability to handle changes in the queries of the target dataset. The intuition is that if a model is robust enough to the queries and documents in the target domain, then the relevance scores should remain stable even after introducing noise to the queries.
To evaluate model robustness, we follow a query perturbation-based approach. Specifically, we issue the original queries to each model on the target domain dataset and record the top k retrieved documents of each query. We then randomly replace some tokens in the original queries with [MASK] tokens, with the proportion of replaced tokens controlled by a hyper-parameter p that ranges from 0 to 1.
Subsequently, we recompute the relevance scores between the altered queries and the originally retrieved documents. By comparing the scores of the original and perturbed queries, we obtain an indication of how well the model can handle variations and noise in query inputs. We quantify the model robustness by calculating the standard deviation of the score changes, with lower standard deviation implying better robustness. 
\section{Experimental Setup}


\subsection{Datasets} 
To evaluate the ability of the predictors to select the most performing model for a target dataset, we employ the BEIR evaluation benchmark~\cite{thakur2021beir}, which has been commonly employed to evaluate the generalisation and zero-shot effectiveness of retrieval models~\cite{gururangan-etal-2020-dont,wang2022gpl,ma2021zero,ren2022thorough}. BEIR contains a heterogeneous set of 18 datasets drawn from 9 text retrieval tasks and domains. We refer the reader to the original work describing BEIR for an in-depth analysis of each dataset and baseline results~\cite{thakur2021beir} -- however, as the complete set of BEIR subsets is not publicly available, we follow the standard practice of selecting a representative subset of the available subsets for our experiments. Specifically, we select 10 out of the 18 available subsets to conduct our experiments. The selected subsets are consistent with prior work on the dataset, allowing for comparability across studies. We highlight that a key finding of that work was that dense retrievers exhibited poor generalization capabilities.

\subsection{Dense Retrieval Models} 
To evaluate our proposed dense retrieval model selection methods, we require a diverse pool of DR models that exhibit varying performance across different BEIR subsets. To this end, we consider DR models that have been submitted to the BEIR leaderboard~\footnote{https://github.com/beir-cellar/beir/wiki/Leaderboard}, including Contriver~\cite{izacardunsupervised}, DistilBERT (TAS-B)~\cite{hofstatter2021efficiently}, ANCE~\cite{xiongapproximate}, DistilBERT v3~\cite{reimers-2019-sentence-bert}, DistilBERT (dot)~\cite{reimers-2019-sentence-bert}, MiniLM-L-12~\cite{reimers-2019-sentence-bert}, MiniLM-L-6~\cite{reimers-2019-sentence-bert}, and DistilBERT v2~\cite{reimers-2019-sentence-bert}. Additionally, we include some advanced DR models that are not on the leaderboard, such as CoCondenser~\cite{gao2021condenser} and SimLM~\cite{wang2022simlm}, as well as a DR model that we trained ourselves (BERT-DPR) by using Tevatron DR training toolkit~\cite{Gao2022TevatronAE}. We obtain all model checkpoints from the Huggingface model hub, as uploaded by the original authors. Notably, all DR models considered in this study are trained on MS MARCO training data~\cite{nguyen2016ms}, and therefore perform zero-shot retrieval on BEIR datasets.
	



\subsection{Evaluation Measures for Dense Retrievers Selection}\label{sec:eval_measures}
 
For evaluating considered model selection methods, we need to first have the ideal model ranking on the target corpus to compare with the rankings generated by the model selection methods. To this end, for each of the considered datasets, we record the effectiveness of each dense retrievers according to several evaluation measures $e$. Specifically we consider nDCG@10, which is the official evaluation measure for the BEIR leaderboard. We report these measures for the Dense Retrievers we consider in our main experiments in Table~\ref{tbl:DR-eval}.

After having the real nDCG@10 scores of methods on BEIR dataset, then 
given a dataset, we identify the best Dense Retriever model $M$ according to the evaluation measure $e$ among the set $\mathcal{M}$ of dense retrievers we consider, i.e. $M = argmax_{\widetilde{M}\in\mathcal{M}}(e(\widetilde{M}))$. For each prediction method $\theta$ we then identify the model $\hat{M}_\theta$ that has been predicted as being the most effective for that dataset. Then, we measure the loss for the evaluation measure $e$ when using $\hat{M}_\theta$, i.e. the model predicted to be the best by the prediction method $\theta$, in place of $M$, i.e. the true best model. We define this loss as:

\begin{equation}
	\Delta_e = e(\hat{M}_\theta) - e(M)
\end{equation}

We also consider the relative loss produced when choosing $\hat{M}_\theta$ in place of $M$: this measure is interesting in that a search engine practitioner may tolerate to select a sub-optimal model up to a certain percentage of loss.

\begin{equation}
	\%\Delta_e = 100 * \frac{e(\hat{M}_\theta) - e(M)}{e(M)}
\end{equation}

In addition the above loss based evaluation method, we also consider the Kendall Tau correlation, which measures the similarity between the rankings generated by the model selection methods and the ranking by true nDCG values. Specifically, the Kendall Tau correlation measures the proportion of pairs of documents that are ranked in the same order by both rankings. A perfect correlation between the rankings of two models results in a Kendall Tau score of 1 in case of positive correlation and -1 in case of negative correlation, while a completely random correlation results in a score of 0.

\section{Results}
Table~\ref{tbl:DR-eval} presents the effectiveness in terms of nDCG@10 of the considered DR models on the source MS MARCO dataset and the target collection BEIR, as well as the in-domain and out-of-domain (zero-shot) ground truth model ranking.  It is evident that the zero-shot ranking differs from the in-domain ranking, highlighting the necessity for a better model selection criterion.
 Furthermore, we provide the effectiveness of our proposed model selection methods in terms of Kendall Tau correlation, $\Delta_e$, and $\%\Delta_e$ in Tables~\ref{tab:tau_ndcg@10}, \ref{tab:delta}, and \ref{tab:percentage_delta}, respectively. These tables allow us to compare the effectiveness of the selected models and provide a comprehensive overview of the investigated methods for model selection.

\begin{table*}[t]
    \resizebox{\textwidth}{!}{
    \centering
    \begin{tabular}{c|c|c|c|c|c|c|c|c|c|c|c}

          \hline
          &  COVID & NF & NQ & HotpotQA & FiQA & ArguAna & DBPedia & SciDocs & SciFact & Quora& \textbf{Avrg}\\
          \hline
          In-Domain Performance (1) & 0.273&   0.455 &  0.782 & 0.709& 0.745& 0.018&   0.6 &  0.636 &   0.745 &  0.273 &0.524
          \\
          Query Similarity (2) & 0.673& 0.127& 0.345& 0.091& 0.2&  -0.164 &  0.055 &  0.091 & 0.018 & -0.091 & 0.135
         \\
         Corpus Similarity (3) & 0.2 & 0.056 & 0.345 & 0.309 &  0.455 & 0.055 & 0.200& 0.382 &  0.309 & 0.164 & 0.247
         \\
         Extracted Doc similarly @100 (4) & 0.273 & -0.018  & 0.491 & 0.200 & 0.345& -0.018 &  0.236& 0.236  & 0.200 & 0.127& 0.207\\
        \hline
         Binary entropy 10 (5) &  0.418 & 0.491& -0.127 & -0.055 & 0.055 & 0.491 & -0.345 & 0.127 &0.309 &  0.236 & 0.16\\
         Binary entropy 1000 (5) &  0.418 & 0.491& -0.127 &  -0.055 & 0.055 & 0.491 & -0.345 & 0.127 &0.309 &  0.236 & 0.16\\
         \hline
         Query Alteration Std p=0.1 (6) &  -0.127 & 0.091& 0.345 &0.309 & 0.273 & 0.127 &  0.164 &  0.309 & 0.2 &0.2&0.189 \\
         Query Alteration Std p=0.2 (6) & -0.164 & 0.055 & 0.345 & 0.273 & 0.273 & 0.164 & 0.164 & 0.2 & 0.2 & 0.236 & 0.175 \\
         Query Alteration Std p=0.3 (6) &-0.236 & -0.018 & 0.236 & 0.164 & 0.2 & 0.091 & 0.164 & 0.127 & 0.091 & 0.164 & 0.098 \\
         \hline

    \end{tabular}
    }
\vspace{5pt}
    \caption{Kendall Tau Correlation value, calculated based on nDCG@10 (the higher, the better)}
    \label{tab:tau_ndcg@10}

\end{table*}

\begin{table*}[t]
    \resizebox{\textwidth}{!}{
    \centering
    \begin{tabular}{c|c|c|c|c|c|c|c|c|c|c|c}

          \hline
          &  TREC-COVID &NFCorpus & NQ & HotpotQA & FiQA & ArguAna & DBPedia & SciDocs & SciFact & Quora& \textbf{Avrg}\\
          \hline
          In-Domain Performance (1) & 0.225 & 0.01 & 0.0 & 0.07 & 0.032 & 0.07 & 0.062 & 0.028 & 0.118 & 0.068 & 0.068
          \\
         Query Similarity (2) & 0.099 & 0.092 & 0.058 & 0.186 & 0.035 & 0.027 & 0.132 & 0.043 & 0.165 & 0.013 & 0.085
         \\
         Corpus Similarity (3) &0.225 & 0.01 & 0.0 & 0.07 & 0.032 & 0.07 & 0.062 & 0.028 & 0.118 & 0.068 & 0.068 \\
Extracted Doc similarly @100 (4) &0.225 & 0.01 & 0.0 & 0.07 & 0.032 & 0.07 & 0.062 & 0.028 & 0.118 & 0.068 & 0.068 \\
         \hline
        Binary entropy @10 (5) &  0.099 & 0.07 & 0.058 & 0.186 & 0.035 & 0.019 & 0.132 & 0.043 & 0.034 & 0.0 & 0.067
        \\
        Binary entropy @1000 (5) & 0.099 & 0.07 & 0.058 & 0.186 & 0.035 & 0.019 & 0.132 & 0.043 & 0.034 & 0.0 & 0.067
        \\
        \hline

        \hline
         Query Alteration Std p=0.1 (6) & 0.225 & 0.01 & 0.0 & 0.07 & 0.032 & 0.07 & 0.062 & 0.028 & 0.118 & 0.068 & 0.068
         \\
        Query Alteration Std p=0.2 (6) & 0.225 & 0.01 & 0.0 & 0.07 & 0.032 & 0.07 & 0.062 & 0.028 & 0.118 & 0.068 & 0.068
        \\
        Query Alteration Std p=0.3 (6)& 0.225 & 0.01 & 0.0 & 0.07 & 0.032 & 0.07 & 0.062 & 0.028 & 0.118 & 0.068 & 0.068 \\
        \hline

         \hline
    \end{tabular}
    }
\vspace{5pt}
    \caption{$\Delta_e$, calculated based on nDCG@10 (the lower, the better)}
    \label{tab:delta}
\end{table*}

\begin{table*}[t]
    \resizebox{\textwidth}{!}{
    \centering
    \begin{tabular}{c|c|c|c|c|c|c|c|c|c|c|c}

        \hline
        &  TREC-COVID &NFCorpus & NQ & HotpotQA & FiQA & ArguAna & DBPedia & SciDocs & SciFact & Quora& \textbf{Avrg}\\
        \hline
        In-Domain Performance (1) &  29.88 & 3.03 & 0.0 & 10.91 & 9.69 & 15.63 & 14.97 & 16.8 & 17.38 & 7.87 & 12.62
        \\
       Query Similarity (2) & 13.12 & 28.16 & 11.52 & 29.23 & 10.5 & 6.09 & 31.9 & 26.25 & 24.44 & 1.46 & 18.27
       \\
       Corpus Similarity (3) &  29.88 & 3.03 & 0.0 & 10.91 & 9.69 & 15.63 & 14.97 & 16.8 & 17.38 & 7.87 & 12.62\\
       Extracted Doc similarly @100  (4) &  29.88 & 3.03 & 0.0 & 10.91 & 9.69 & 15.63 & 14.97 & 16.8 & 17.38 & 7.87 & 12.62\\

       \hline
      Binary entropy @10 (5) & 13.12 & 21.36 & 11.52 & 29.23 & 10.5 & 4.29 & 31.9 & 26.25 & 5.03 & 0.0 & 15.32
      \\
      Binary entropy @1000 (5) & 13.12 & 21.36 & 11.52 & 29.23 & 10.5 & 4.29 & 31.9 & 26.25 & 5.03 & 0.0 & 15.32
      \\
      \hline

      \hline
       Query Alteration Std p=0.1 (6) & 29.88 & 3.03 & 0.0 & 10.91 & 9.69 & 15.63 & 14.97 & 16.8 & 17.38 & 7.87 & 12.62
       \\
      Query Alteration Std p=0.2 (6) & 29.88 & 3.03 & 0.0 & 10.91 & 9.69 & 15.63 & 14.97 & 16.8 & 17.38 & 7.87 & 12.62
      \\
      Query Alteration Std p=0.3 (6)& 29.88 & 3.03 & 0.0 & 10.91 & 9.69 & 15.63 & 14.97 & 16.8 & 17.38 & 7.87 & 12.62 \\
      \hline

       \hline
  \end{tabular}
    }
\vspace{6pt}
    \caption{$\% \Delta_e$, calculated based on nDCG@10  (the lower, the better) }
    \label{tab:percentage_delta}
\end{table*}

\subsection{Desired ranking}
To gain an understanding of the desired model ranking, we first evaluate the actual effectiveness of the considered DR models on the BEIR dataset. The results are presented in Table~\ref{tbl:DR-eval}. While Contriever may not perform the best on the in-domain corpus (i.e. the dataset that is aligned to the training data), it outperforms other DRs on most of the zero-shot datasets, ranking at number one in terms of average zero-shot effectiveness. CoCondense and SimLM, on the other hand, exhibit better in-domain effectiveness than Contriever, but only achieve the best zero-shot effectiveness on NQ and COVID, respectively, ranking at four and two on average in terms of zero-shot effectiveness. Similarly, the effectiveness of all other models varies on the zero-shot datasets, irrespective of their effectiveness on the in-domain data.

It is important to keep in mind that our objective is to rank the DR models on the target dataset as closely as possible to the true ranking of the models. Therefore, the naive baseline of ranking models solely based on their in-domain effectiveness is evidently inadequate, which is quantified in Tables~\ref{tab:tau_ndcg@10}, \ref{tab:delta} and~\ref{tab:percentage_delta} (first row of each table).

We next analyze how each of the methods we considered for model selection perform, compared to the naive baseline of ranking models solely based on their in-domain effectiveness.

\subsection{Query Similarity}

The Query Similarity method identifies ANCE as the top-ranked model across all zero-shot datasets. This may be because most of ANCE's top scores have values close to 1, unlike the other DR models. 
CoCondenser is always ranked as the second best model, followed by BERT-DPR in third place. However, it should be noted that predicting BERT-DPR to be ranked high is not desirable, as its actual rank position based on the ground-truth is usually near the bottom of the list, regardless of the zero-shot dataset, as shown in Table~\ref{tbl:DR-eval}.
Overall, Query Similarity displays weak average correlation with the true rankings of DR models (Table~\ref{tab:tau_ndcg@10}), and is the worst method in terms of $\Delta_e$ and $\%\Delta_e$.


\subsection{Corpus Similarity}
According to the Corpus Similarity method, SimLM is consistently ranked as the best model across all datasets, followed by Contriever as the second-best model. The third-place ranking varies between ANCE and MiniLM-L-12 depending on the zero-shot dataset. The bottom-ranking models are almost identical across all datasets. 
Notably, this method provides the best Kendall Tau correlation among all considered methods, but is inferior to the In-Domain performance baseline. Additionally, unlike most other unsupervised methods, the Fréchet distance, which defines the corpus similarity,  consistently exhibits a negative correlation with the accuracy across all target datasets. This characteristic  makes Corpus Similarity the most reliable measure for unsupervised model selection among considered methods.

\subsection{Extracted Document Similarity}
The ranking of the top three DRs according to the Extracted Document Similarity method is consistent across all datasets, with Contriever securing the first place, followed by ANCE and MiniLM-12. Although this method is slightly less effective than corpus similarity, it still yields the best $\Delta_e$ and $\%\Delta_e$ values, matching those of the baseline as well as those of some of the other methods.

\subsection{Binary Entropy}
Binary Entropy is the only method that selects different models across different zero-shot datasets: either DistilBERT (TAS-B) or ANCE. According to Table \ref*{tab:delta} the cut-off of the documents per query does not matter as it does not affect the ranking of models. Binary Entropy provides the best ranking of DRs for the Arguana and NF datasets ($\tau=0.491$), even outperforming the In-Domain performance method on these datasets.
  However, the biggest drawback of Binary Entropy is that it does not have a consistent correlation pattern across datasets: for example, it shows moderate negative correlation for DBPedia with $\tau=-0.345$, while moderate positive correlation of the two datasets for which it performs best, Arguana and NF. The analysis of $\Delta_e$ and $\%\Delta_e$ values paint a similar picture. We discuss the origins of this inconsistency in the next section.

\subsection{Query Alteration}
Across all datasets, Query Alteration indicates as the best model SimLM, followed by DistillBert. The method  displays the lowest average $\%\Delta_e$, at par with the baseline; the individual values of $\%\Delta_e$ across the zero-shot datasets are often the best recorded, except in the case of the TREC-COVID dataset. We also note that the $\%\Delta_e$ of this method matches exactly the one of the In-Domain performance baseline. Note that the probability $p$ of replaced tokens affects the effectiveness of the predictions, with the best $p$ being 0.1.

\section{Outlook}

We have examined the effectiveness of adapting methods of unsupervised performance evaluation with the presence of domain shift, proposed in the area of computer vision and machine learning, to the problem of choosing which dense retriever model to use when searching on a new collection for which no labels are available, i.e. in a zero-shot setting.

Our first finding is that the in-domain effectiveness of the dense retrievers, trained on a large corpus like MS MARCO, represents a strong indicator of model generalizability on the target domains, despite the existence of domain gap. 
However, we note that the problem of dense retrieval selection, i.e. which dense retriever model to select in a new domain, remains an important and unresolved issue. Specifically, we observe that the straightforward in-domain-based model selection method fails to choose the best model across all the datasets examined. Furthermore, for certain datasets, the ranking of models based on this method is significantly flawed. The methods we have investigated and that we have adapted from computer vision and machine learning can only do as well as this simple baseline (though some do notably worse).
These findings highlight the need for further research to improve the effectiveness of model selection techniques to make feasible the practical application of dense retrieval in domain-specific tasks characterized by low labeled data availability.

In this work, we focus on a realistic model selection scenario in information retrieval, which differs from the artificially-created setups investigated in prior work in machine learning (see Challenge 1 in Section~\ref{subsection:challenges}). Due to the challenges in adapting model selection methods to the information retrieval setup, we primarily considered two broad groups of methods: uncertainty-based and activation-based. 

\textbf{Uncertainty-based methods.} In the field of machine learning, entropy-based methods are often used to measure the level of uncertainty of a model. These methods analyze the score distribution produced by the model and use it to assess the model's level of confidence in its predictions.

In classification tasks, the concept of uncertainty is naturally linked to the score distribution since there is only one correct class per input sample. When a model predicts two classes with similar probabilities, it indicates uncertainty about the sample's classification.

In information retrieval tasks, the requirement for an ideal ranker is to maximize the binary entropy of relevant document predictions and minimize the binary entropy of irrelevant ones, as defined by~\citet{10.1145/1076034.1076042}. This presents a challenge since the number of relevant documents in an unjudged collection is unknown. As a result, the connection between score distribution and model uncertainty in ranking tasks is not intuitive. 
For instance, a retriever that produces a steep score distribution with a top-ranked document having a significantly higher score than other retrieved documents will have a small entropy score. On the other hand, a retriever with a more gradually-declining score distribution will have a larger entropy score. However, having a large entropy of predictions does not necessarily imply that the model is uncertain. It may instead reflect that the model predicts more of the top-k retrieved documents to be equally relevant. In addition, the nature of the dataset itself could highly influence the score distributions. Large datasets with many documents that are relevant to a query will likely result in rankings with low entropy; while small datasets with very different documents and with just a handful of documents that are close to the query will likely result in rankings with a high entropy. In this example, however, the case with higher entropy does not indicate the model is more certain about its predictions.


Another way to define the uncertainty of the model is to evaluate its robustness to query perturbation. The intuition is that if the model is confident in its predictions, it should not be affected by a slight perturbation of the query. Consequently, network robustness can be used as a quality measure of the model, and employed as an unsupervised criterion for model selection. Indeed, according to our results, this approach provides competitive model ranking, with an average Tau Correlation of 0.189. Despite this, the correlation is very weak and the method falls noticeably behind the top-performing baseline, indicating that there is still ample opportunity for improvement, which we recommend the information retrieval community should explore in future research.

The issue of uncertainty estimation in information retrieval is a significant yet largely under-explored problem, as noted in several prior studies~\cite{turtle1997uncertainty, crestani1998information, collins2007estimation}. This is particularly true for rankers that rely on pre-trained language models~\cite{lesota2021modern}. While some efforts have been made to leverage uncertainty in relevance estimation for traditional keyword-based best-match models like BM25 and language models, the current approaches are based on assumptions and heuristics that use similarities or covariance between term occurrences~\cite{zhu2009risky, wang2009portfolio, zuccon2009quantum}, follow the Dirichlet distribution~\cite{wang2009portfolio}, or calculate score distributions by resampling query terms~\cite{collins2007estimation}. Recently, researchers have attempted to model uncertainty for neural rankers, such as with the Transformer Pointer Generator Network (T-PGN) model~\cite{lesota2021modern} 
However, these models are not easily adaptable to the dense retrieval ranker architectures that we have examined in this paper. More effective and meaningful estimations of uncertainty might be a promising future direction also in our context.

\textbf{Activation-based methods.}
Activation-based methods approximate the domain gap between the source and target distributions via the proximity of its hidden representations. The underlying idea is that a model with good generalization capabilities should produce similar encodings for both datasets.
The advantage of this method if adapted to information retrieval is that it is not affected by variations in network architectures or scoring functions used, which enables the comparison of models within the constraints of the information retrieval setup (i.e. across different pre-trained language models and architectures). This makes it easier to identify models that perform well on both source and target datasets, without being biased towards specific network architectures or scoring functions. 

In our experiments, we used the Fréchet distance between the activations of the network for the source and target datasets (Method 3, Corpus Similarity).  This approach, highly effective for example in computer vision tasks, in our experiments selects the same best model for all the datasets, similarly to the in-domain performance-based method, but produces an inferior model ranking (see Table \ref*{tab:tau_ndcg@10}). To handle the high dimensionality of the activation spaces, the Fréchet-based approximation assumes that they conform to a normal distribution~\cite{Chong2019EffectivelyUF}, which may not hold for NLP-induced hidden representations. We believe that designing NLP-oriented distance approximations between activation spaces is a promising future direction for more accurate unsupervised model selection, especially in the context of selecting dense retrieval models.  

\textbf{Weight- and Performance-based methods.} The last direction we would like to mention is inspired the recent work by \citet{khramtsova} and \citet{Deng2021WhatDR} which involves fine-tuning the network on the target dataset with the unsupervised loss (e.g., it can be a masked token prediction task) and evaluating the degree of the network change or the performance on the unsupervised task. Despite being computationally expensive, these approaches have been shown very effective in general machine learning model selection, and can be further explored in the information retrieval context.

\section{Conclusion}

This paper proposes a novel research direction for zero-shot dense retrieval. While traditional information retrieval research in this area concentrates on developing universal domain-agnostic DR models, our work shifts the focus towards developing a method to rank and select pre-trained state-of-the-art DR models that are best suited for a specific target domain corpus.
 We acknowledge that the proposed direction does not contradict traditional research on training zero-shot DR models, but rather complements it. As newly developed DR models are likely to have varying effects on different domains, selecting the best model is still beneficial.
To explore this research direction, we adapt various methods from computer vision and machine learning, along with some approaches designed for IR. We outline our reasoning and challenges with the investigated approaches and present empirical results on a popular zero-shot benchmark dataset. Our findings shed light on future research avenues within this research direction.
We believe that an effective method for selecting a good DR model can provide a principled way for search engine developers to identify the most suitable model for their application, ultimately enhancing user experience.

\bibliographystyle{ACM-Reference-Format}
\bibliography{Bib/bibliography}


\begin{thebibliography}{55}


\ifx \showCODEN    \undefined \def \showCODEN     #1{\unskip}     \fi
\ifx \showDOI      \undefined \def \showDOI       #1{#1}\fi
\ifx \showISBNx    \undefined \def \showISBNx     #1{\unskip}     \fi
\ifx \showISBNxiii \undefined \def \showISBNxiii  #1{\unskip}     \fi
\ifx \showISSN     \undefined \def \showISSN      #1{\unskip}     \fi
\ifx \showLCCN     \undefined \def \showLCCN      #1{\unskip}     \fi
\ifx \shownote     \undefined \def \shownote      #1{#1}          \fi
\ifx \showarticletitle \undefined \def \showarticletitle #1{#1}   \fi
\ifx \showURL      \undefined \def \showURL       {\relax}        \fi
\providecommand\bibfield[2]{#2}
\providecommand\bibinfo[2]{#2}
\providecommand\natexlab[1]{#1}
\providecommand\showeprint[2][]{arXiv:#2}

\bibitem[\protect\citeauthoryear{Aslam, Yilmaz, and Pavlu}{Aslam
  et~al\mbox{.}}{2005}]%
        {10.1145/1076034.1076042}
\bibfield{author}{\bibinfo{person}{Javed~A. Aslam}, \bibinfo{person}{Emine
  Yilmaz}, {and} \bibinfo{person}{Virgiliu Pavlu}.}
  \bibinfo{year}{2005}\natexlab{}.
\newblock \showarticletitle{The Maximum Entropy Method for Analyzing Retrieval
  Measures}. In \bibinfo{booktitle}{\emph{Proceedings of the 28th Annual
  International ACM SIGIR Conference on Research and Development in Information
  Retrieval}} (Salvador, Brazil) \emph{(\bibinfo{series}{SIGIR '05})}.
  \bibinfo{publisher}{Association for Computing Machinery},
  \bibinfo{address}{New York, NY, USA}, \bibinfo{pages}{27--34}.
\newblock
\showISBNx{1595930345}
\urldef\tempurl%
\url{https://doi.org/10.1145/1076034.1076042}
\showDOI{\tempurl}


\bibitem[\protect\citeauthoryear{Birdal, Lou, Guibas, and cSimcsekli}{Birdal
  et~al\mbox{.}}{2021}]%
        {Birdal2021IntrinsicDP}
\bibfield{author}{\bibinfo{person}{Tolga Birdal}, \bibinfo{person}{Aaron Lou},
  \bibinfo{person}{Leonidas~J. Guibas}, {and} \bibinfo{person}{Umut
  cSimcsekli}.} \bibinfo{year}{2021}\natexlab{}.
\newblock \showarticletitle{Intrinsic Dimension, Persistent Homology and
  Generalization in Neural Networks}. In \bibinfo{booktitle}{\emph{NeurIPS}}.
\newblock


\bibitem[\protect\citeauthoryear{Bonifacio, Abonizio, Fadaee, and
  Nogueira}{Bonifacio et~al\mbox{.}}{2022}]%
        {bonifacio2022inpars}
\bibfield{author}{\bibinfo{person}{Luiz Bonifacio}, \bibinfo{person}{Hugo
  Abonizio}, \bibinfo{person}{Marzieh Fadaee}, {and} \bibinfo{person}{Rodrigo
  Nogueira}.} \bibinfo{year}{2022}\natexlab{}.
\newblock \showarticletitle{Inpars: Data augmentation for information retrieval
  using large language models}.
\newblock \bibinfo{journal}{\emph{arXiv preprint arXiv:2202.05144}}
  (\bibinfo{year}{2022}).
\newblock


\bibitem[\protect\citeauthoryear{Carmel and Kurland}{Carmel and
  Kurland}{2012}]%
        {carmel2012query}
\bibfield{author}{\bibinfo{person}{David Carmel} {and} \bibinfo{person}{Oren
  Kurland}.} \bibinfo{year}{2012}\natexlab{}.
\newblock \showarticletitle{Query performance prediction for ir}. In
  \bibinfo{booktitle}{\emph{Proceedings of the 35th international ACM SIGIR
  conference on Research and development in information retrieval}}.
  \bibinfo{pages}{1196--1197}.
\newblock


\bibitem[\protect\citeauthoryear{Chen, Liu, Avci, Wu, Liang, and Jha}{Chen
  et~al\mbox{.}}{2021}]%
        {Chen2021DetectingEA}
\bibfield{author}{\bibinfo{person}{Jiefeng Chen}, \bibinfo{person}{Frederick
  Liu}, \bibinfo{person}{Besim Avci}, \bibinfo{person}{Xi Wu},
  \bibinfo{person}{Yingyu Liang}, {and} \bibinfo{person}{Somesh Jha}.}
  \bibinfo{year}{2021}\natexlab{}.
\newblock \showarticletitle{Detecting Errors and Estimating Accuracy on
  Unlabeled Data with Self-training Ensembles}. In
  \bibinfo{booktitle}{\emph{Advances in Neural Information Processing
  Systems}}.
\newblock


\bibitem[\protect\citeauthoryear{Chen, Zhang, Lu, Bendersky, and Najork}{Chen
  et~al\mbox{.}}{2022}]%
        {Chen2022OutofDomainST}
\bibfield{author}{\bibinfo{person}{Tao Chen}, \bibinfo{person}{Mingyang Zhang},
  \bibinfo{person}{Jing Lu}, \bibinfo{person}{Michael Bendersky}, {and}
  \bibinfo{person}{Marc Najork}.} \bibinfo{year}{2022}\natexlab{}.
\newblock \showarticletitle{Out-of-Domain Semantics to the Rescue! Zero-Shot
  Hybrid Retrieval Models}. In \bibinfo{booktitle}{\emph{European Conference on
  Information Retrieval}}.
\newblock


\bibitem[\protect\citeauthoryear{Chong and Forsyth}{Chong and Forsyth}{2020a}]%
        {DOWSON1982450}
\bibfield{author}{\bibinfo{person}{Min~Jin Chong} {and}
  \bibinfo{person}{David~A. Forsyth}.} \bibinfo{year}{2020}\natexlab{a}.
\newblock \showarticletitle{Effectively Unbiased FID and Inception Score and
  Where to Find Them}.
\newblock  (\bibinfo{year}{2020}), \bibinfo{pages}{6069--6078}.
\newblock


\bibitem[\protect\citeauthoryear{Chong and Forsyth}{Chong and Forsyth}{2020b}]%
        {Chong2019EffectivelyUF}
\bibfield{author}{\bibinfo{person}{Min~Jin Chong} {and}
  \bibinfo{person}{David~A. Forsyth}.} \bibinfo{year}{2020}\natexlab{b}.
\newblock \showarticletitle{Effectively Unbiased FID and Inception Score and
  Where to Find Them}.
\newblock  (\bibinfo{year}{2020}), \bibinfo{pages}{6069--6078}.
\newblock


\bibitem[\protect\citeauthoryear{Chuang, Torralba, and Jegelka}{Chuang
  et~al\mbox{.}}{2020}]%
        {chuang2020estimating}
\bibfield{author}{\bibinfo{person}{Ching-Yao Chuang}, \bibinfo{person}{Antonio
  Torralba}, {and} \bibinfo{person}{Stefanie Jegelka}.}
  \bibinfo{year}{2020}\natexlab{}.
\newblock \showarticletitle{Estimating Generalization under Distribution Shifts
  via Domain-Invariant Representations}.
\newblock \bibinfo{journal}{\emph{International Conference of Machine Learning
  (ICML)}} (\bibinfo{year}{2020}).
\newblock


\bibitem[\protect\citeauthoryear{Collins-Thompson and Callan}{Collins-Thompson
  and Callan}{2007}]%
        {collins2007estimation}
\bibfield{author}{\bibinfo{person}{Kevyn Collins-Thompson} {and}
  \bibinfo{person}{Jamie Callan}.} \bibinfo{year}{2007}\natexlab{}.
\newblock \showarticletitle{Estimation and use of uncertainty in
  pseudo-relevance feedback}. In \bibinfo{booktitle}{\emph{Proceedings of the
  30th annual international ACM SIGIR conference on Research and development in
  information retrieval}}. \bibinfo{pages}{303--310}.
\newblock


\bibitem[\protect\citeauthoryear{Corneanu, Madadi, Escalera, and
  Martinez}{Corneanu et~al\mbox{.}}{2019}]%
        {Corneanu2019}
\bibfield{author}{\bibinfo{person}{Ciprian~A. Corneanu},
  \bibinfo{person}{Meysam Madadi}, \bibinfo{person}{Sergio Escalera}, {and}
  \bibinfo{person}{Aleix~M. Martinez}.} \bibinfo{year}{2019}\natexlab{}.
\newblock \showarticletitle{What Does It Mean to Learn in Deep Networks? And,
  How Does One Detect Adversarial Attacks?}. In
  \bibinfo{booktitle}{\emph{IEEE/CVF Conference on Computer Vision and Pattern
  Recognition}}.
\newblock


\bibitem[\protect\citeauthoryear{Crestani, Lalmas, and van Rijsbergen}{Crestani
  et~al\mbox{.}}{1998}]%
        {crestani1998information}
\bibfield{author}{\bibinfo{person}{Fabio Crestani}, \bibinfo{person}{Mounia
  Lalmas}, {and} \bibinfo{person}{Cornelis~Joost van Rijsbergen}.}
  \bibinfo{year}{1998}\natexlab{}.
\newblock \bibinfo{booktitle}{\emph{Information retrieval: Uncertainty and
  logics: Uncertainty and logics: Advanced models for the representation and
  retrieval of information}}. Vol.~\bibinfo{volume}{4}.
\newblock \bibinfo{publisher}{Springer Science \& Business Media}.
\newblock


\bibitem[\protect\citeauthoryear{Deng, Gould, and Zheng}{Deng
  et~al\mbox{.}}{2021}]%
        {Deng2021WhatDR}
\bibfield{author}{\bibinfo{person}{Weijian Deng}, \bibinfo{person}{Stephen
  Gould}, {and} \bibinfo{person}{Liang Zheng}.}
  \bibinfo{year}{2021}\natexlab{}.
\newblock \showarticletitle{What Does Rotation Prediction Tell Us about
  Classifier Accuracy under Varying Testing Environments?}. In
  \bibinfo{booktitle}{\emph{International Conference of Machine Learning
  (ICML)}}.
\newblock


\bibitem[\protect\citeauthoryear{Deng and Zheng}{Deng and Zheng}{2021}]%
        {deng2020labels}
\bibfield{author}{\bibinfo{person}{Weijian Deng} {and} \bibinfo{person}{Liang
  Zheng}.} \bibinfo{year}{2021}\natexlab{}.
\newblock \showarticletitle{Are Labels Always Necessary for Classifier Accuracy
  Evaluation?}. In \bibinfo{booktitle}{\emph{IEEE Conference on Computer Vision
  and Pattern Recognition (CVPR)}}.
\newblock


\bibitem[\protect\citeauthoryear{Gao and Callan}{Gao and Callan}{2021}]%
        {gao2021condenser}
\bibfield{author}{\bibinfo{person}{Luyu Gao} {and} \bibinfo{person}{Jamie
  Callan}.} \bibinfo{year}{2021}\natexlab{}.
\newblock \showarticletitle{Condenser: a Pre-training Architecture for Dense
  Retrieval}. In \bibinfo{booktitle}{\emph{Proceedings of the 2021 Conference
  on Empirical Methods in Natural Language Processing}}.
  \bibinfo{pages}{981--993}.
\newblock


\bibitem[\protect\citeauthoryear{Gao, Ma, Lin, and Callan}{Gao
  et~al\mbox{.}}{2022}]%
        {Gao2022TevatronAE}
\bibfield{author}{\bibinfo{person}{Luyu Gao}, \bibinfo{person}{Xueguang Ma},
  \bibinfo{person}{Jimmy~J. Lin}, {and} \bibinfo{person}{Jamie Callan}.}
  \bibinfo{year}{2022}\natexlab{}.
\newblock \showarticletitle{Tevatron: An Efficient and Flexible Toolkit for
  Dense Retrieval}.
\newblock \bibinfo{journal}{\emph{ArXiv}}  \bibinfo{volume}{abs/2203.05765}
  (\bibinfo{year}{2022}).
\newblock


\bibitem[\protect\citeauthoryear{Garg, Balakrishnan, Lipton, Neyshabur, and
  Sedghi}{Garg et~al\mbox{.}}{2022}]%
        {ATC}
\bibfield{author}{\bibinfo{person}{Saurabh Garg}, \bibinfo{person}{Sivaraman
  Balakrishnan}, \bibinfo{person}{Zachary~Chase Lipton},
  \bibinfo{person}{Behnam Neyshabur}, {and} \bibinfo{person}{Hanie Sedghi}.}
  \bibinfo{year}{2022}\natexlab{}.
\newblock \showarticletitle{Leveraging Unlabeled Data to Predict
  Out-of-Distribution Performance}. In \bibinfo{booktitle}{\emph{International
  Conference on Learning Representations, {ICLR}}}.
\newblock
\urldef\tempurl%
\url{https://arxiv.org/abs/2201.04234}
\showURL{%
\tempurl}


\bibitem[\protect\citeauthoryear{Guillory, Shankar, Ebrahimi, Darrell, and
  Schmidt}{Guillory et~al\mbox{.}}{2021}]%
        {Guillory2021PredictingWC}
\bibfield{author}{\bibinfo{person}{Devin Guillory}, \bibinfo{person}{Vaishaal
  Shankar}, \bibinfo{person}{Sayna Ebrahimi}, \bibinfo{person}{Trevor Darrell},
  {and} \bibinfo{person}{Ludwig Schmidt}.} \bibinfo{year}{2021}\natexlab{}.
\newblock \showarticletitle{Predicting with Confidence on Unseen
  Distributions}.
\newblock \bibinfo{journal}{\emph{IEEE/CVF International Conference on Computer
  Vision (ICCV)}} (\bibinfo{year}{2021}), \bibinfo{pages}{1114--1124}.
\newblock


\bibitem[\protect\citeauthoryear{Gururangan, Marasovi{\'c}, Swayamdipta, Lo,
  Beltagy, Downey, and Smith}{Gururangan et~al\mbox{.}}{2020}]%
        {gururangan-etal-2020-dont}
\bibfield{author}{\bibinfo{person}{Suchin Gururangan}, \bibinfo{person}{Ana
  Marasovi{\'c}}, \bibinfo{person}{Swabha Swayamdipta}, \bibinfo{person}{Kyle
  Lo}, \bibinfo{person}{Iz Beltagy}, \bibinfo{person}{Doug Downey}, {and}
  \bibinfo{person}{Noah~A. Smith}.} \bibinfo{year}{2020}\natexlab{}.
\newblock \showarticletitle{Don{'}t Stop Pretraining: Adapt Language Models to
  Domains and Tasks}. In \bibinfo{booktitle}{\emph{Proceedings of the 58th
  Annual Meeting of the Association for Computational Linguistics}}.
  \bibinfo{publisher}{Association for Computational Linguistics},
  \bibinfo{address}{Online}, \bibinfo{pages}{8342--8360}.
\newblock
\urldef\tempurl%
\url{https://aclanthology.org/2020.acl-main.740}
\showURL{%
\tempurl}


\bibitem[\protect\citeauthoryear{Hauff, Hiemstra, and de~Jong}{Hauff
  et~al\mbox{.}}{2008}]%
        {hauff2008survey}
\bibfield{author}{\bibinfo{person}{Claudia Hauff}, \bibinfo{person}{Djoerd
  Hiemstra}, {and} \bibinfo{person}{Franciska de Jong}.}
  \bibinfo{year}{2008}\natexlab{}.
\newblock \showarticletitle{A survey of pre-retrieval query performance
  predictors}. In \bibinfo{booktitle}{\emph{Proceedings of the 17th ACM
  conference on Information and knowledge management}}.
  \bibinfo{pages}{1419--1420}.
\newblock


\bibitem[\protect\citeauthoryear{He and Ounis}{He and Ounis}{2006}]%
        {he2006query}
\bibfield{author}{\bibinfo{person}{Ben He} {and} \bibinfo{person}{Iadh Ounis}.}
  \bibinfo{year}{2006}\natexlab{}.
\newblock \showarticletitle{Query performance prediction}.
\newblock \bibinfo{journal}{\emph{Information Systems}} \bibinfo{volume}{31},
  \bibinfo{number}{7} (\bibinfo{year}{2006}), \bibinfo{pages}{585--594}.
\newblock


\bibitem[\protect\citeauthoryear{Hendrycks and Gimpel}{Hendrycks and
  Gimpel}{2017}]%
        {hendrycks17baseline}
\bibfield{author}{\bibinfo{person}{Dan Hendrycks} {and} \bibinfo{person}{Kevin
  Gimpel}.} \bibinfo{year}{2017}\natexlab{}.
\newblock \showarticletitle{A Baseline for Detecting Misclassified and
  Out-of-Distribution Examples in Neural Networks}.
\newblock \bibinfo{journal}{\emph{Proceedings of International Conference on
  Learning Representations}} (\bibinfo{year}{2017}).
\newblock


\bibitem[\protect\citeauthoryear{Hofst{\"a}tter, Lin, Yang, Lin, and
  Hanbury}{Hofst{\"a}tter et~al\mbox{.}}{2021}]%
        {hofstatter2021efficiently}
\bibfield{author}{\bibinfo{person}{Sebastian Hofst{\"a}tter},
  \bibinfo{person}{Sheng-Chieh Lin}, \bibinfo{person}{Jheng-Hong Yang},
  \bibinfo{person}{Jimmy Lin}, {and} \bibinfo{person}{Allan Hanbury}.}
  \bibinfo{year}{2021}\natexlab{}.
\newblock \showarticletitle{Efficiently teaching an effective dense retriever
  with balanced topic aware sampling}. In \bibinfo{booktitle}{\emph{Proceedings
  of the 44th International ACM SIGIR Conference on Research and Development in
  Information Retrieval}}. \bibinfo{pages}{113--122}.
\newblock


\bibitem[\protect\citeauthoryear{Izacard, Caron, Hosseini, Riedel, Bojanowski,
  Joulin, and Grave}{Izacard et~al\mbox{.}}{[n.d.]}]%
        {izacardunsupervised}
\bibfield{author}{\bibinfo{person}{Gautier Izacard}, \bibinfo{person}{Mathilde
  Caron}, \bibinfo{person}{Lucas Hosseini}, \bibinfo{person}{Sebastian Riedel},
  \bibinfo{person}{Piotr Bojanowski}, \bibinfo{person}{Armand Joulin}, {and}
  \bibinfo{person}{Edouard Grave}.} \bibinfo{year}{[n.d.]}\natexlab{}.
\newblock \showarticletitle{Unsupervised Dense Information Retrieval with
  Contrastive Learning}.
\newblock \bibinfo{journal}{\emph{Transactions on Machine Learning Research}}
  (\bibinfo{year}{[n.\,d.]}).
\newblock


\bibitem[\protect\citeauthoryear{Jeronymo, Bonifacio, Abonizio, Fadaee, Lotufo,
  Zavrel, and Nogueira}{Jeronymo et~al\mbox{.}}{2023}]%
        {jeronymo2023inpars}
\bibfield{author}{\bibinfo{person}{Vitor Jeronymo}, \bibinfo{person}{Luiz
  Bonifacio}, \bibinfo{person}{Hugo Abonizio}, \bibinfo{person}{Marzieh
  Fadaee}, \bibinfo{person}{Roberto Lotufo}, \bibinfo{person}{Jakub Zavrel},
  {and} \bibinfo{person}{Rodrigo Nogueira}.} \bibinfo{year}{2023}\natexlab{}.
\newblock \showarticletitle{InPars-v2: Large Language Models as Efficient
  Dataset Generators for Information Retrieval}.
\newblock \bibinfo{journal}{\emph{arXiv preprint arXiv:2301.01820}}
  (\bibinfo{year}{2023}).
\newblock


\bibitem[\protect\citeauthoryear{Jiang, Krishnan, Mobahi, and Bengio}{Jiang
  et~al\mbox{.}}{2019}]%
        {jiang2018predicting}
\bibfield{author}{\bibinfo{person}{Yiding Jiang}, \bibinfo{person}{Dilip
  Krishnan}, \bibinfo{person}{Hossein Mobahi}, {and} \bibinfo{person}{Samy
  Bengio}.} \bibinfo{year}{2019}\natexlab{}.
\newblock \showarticletitle{Predicting the Generalization Gap in Deep Networks
  with Margin Distributions}. In \bibinfo{booktitle}{\emph{International
  Conference on Learning Representations}}.
\newblock
\urldef\tempurl%
\url{https://openreview.net/forum?id=HJlQfnCqKX}
\showURL{%
\tempurl}


\bibitem[\protect\citeauthoryear{Jiang, Nagarajan, Baek, and Kolter}{Jiang
  et~al\mbox{.}}{2022}]%
        {Jiang2022AssessingGO}
\bibfield{author}{\bibinfo{person}{Yiding Jiang}, \bibinfo{person}{Vaishnavh
  Nagarajan}, \bibinfo{person}{Christina Baek}, {and} \bibinfo{person}{J.~Zico
  Kolter}.} \bibinfo{year}{2022}\natexlab{}.
\newblock \showarticletitle{Assessing Generalization of SGD via Disagreement}.
\newblock \bibinfo{journal}{\emph{ArXiv}}  \bibinfo{volume}{abs/2106.13799}
  (\bibinfo{year}{2022}).
\newblock


\bibitem[\protect\citeauthoryear{Khratmtsova, Zuccon, Xi, and
  Baktashmotlagh}{Khratmtsova et~al\mbox{.}}{2023}]%
        {khramtsova}
\bibfield{author}{\bibinfo{person}{Ekaterina Khratmtsova},
  \bibinfo{person}{Guido Zuccon}, \bibinfo{person}{Wang Xi}, {and}
  \bibinfo{person}{Mahsa Baktashmotlagh}.} \bibinfo{year}{2023}\natexlab{}.
\newblock \showarticletitle{Weight-Based Performance Estimation for Diverse
  Domains}.
\newblock \bibinfo{journal}{\emph{arXiv preprint}} (\bibinfo{year}{2023}).
\newblock


\bibitem[\protect\citeauthoryear{Lacombe, Ike, and Umeda}{Lacombe
  et~al\mbox{.}}{2021}]%
        {Lacombe2021TopologicalUM}
\bibfield{author}{\bibinfo{person}{Th{\'e}o Lacombe}, \bibinfo{person}{Yuichi
  Ike}, {and} \bibinfo{person}{Yuhei Umeda}.} \bibinfo{year}{2021}\natexlab{}.
\newblock \showarticletitle{Topological Uncertainty: Monitoring trained neural
  networks through persistence of activation graphs}. In
  \bibinfo{booktitle}{\emph{IJCAI}}.
\newblock


\bibitem[\protect\citeauthoryear{Lesota, Rekabsaz, Cohen, Grasserbauer,
  Eickhoff, and Schedl}{Lesota et~al\mbox{.}}{2021}]%
        {lesota2021modern}
\bibfield{author}{\bibinfo{person}{Oleg Lesota}, \bibinfo{person}{Navid
  Rekabsaz}, \bibinfo{person}{Daniel Cohen}, \bibinfo{person}{Klaus~Antonius
  Grasserbauer}, \bibinfo{person}{Carsten Eickhoff}, {and}
  \bibinfo{person}{Markus Schedl}.} \bibinfo{year}{2021}\natexlab{}.
\newblock \showarticletitle{A modern perspective on query likelihood with deep
  generative retrieval models}. In \bibinfo{booktitle}{\emph{Proceedings of the
  2021 ACM SIGIR International Conference on Theory of Information Retrieval}}.
  \bibinfo{pages}{185--195}.
\newblock


\bibitem[\protect\citeauthoryear{Lin, Asai, Li, Oguz, Lin, Mehdad, Yih, and
  Chen}{Lin et~al\mbox{.}}{2023}]%
        {lin2023train}
\bibfield{author}{\bibinfo{person}{Sheng-Chieh Lin}, \bibinfo{person}{Akari
  Asai}, \bibinfo{person}{Minghan Li}, \bibinfo{person}{Barlas Oguz},
  \bibinfo{person}{Jimmy Lin}, \bibinfo{person}{Yashar Mehdad},
  \bibinfo{person}{Wen-tau Yih}, {and} \bibinfo{person}{Xilun Chen}.}
  \bibinfo{year}{2023}\natexlab{}.
\newblock \showarticletitle{How to Train Your DRAGON: Diverse Augmentation
  Towards Generalizable Dense Retrieval}.
\newblock \bibinfo{journal}{\emph{arXiv preprint arXiv:2302.07452}}
  (\bibinfo{year}{2023}).
\newblock


\bibitem[\protect\citeauthoryear{Ma, Korotkov, Yang, Hall, and McDonald}{Ma
  et~al\mbox{.}}{2021}]%
        {ma2021zero}
\bibfield{author}{\bibinfo{person}{Ji Ma}, \bibinfo{person}{Ivan Korotkov},
  \bibinfo{person}{Yinfei Yang}, \bibinfo{person}{Keith Hall}, {and}
  \bibinfo{person}{Ryan McDonald}.} \bibinfo{year}{2021}\natexlab{}.
\newblock \showarticletitle{Zero-shot Neural Passage Retrieval via
  Domain-targeted Synthetic Question Generation}. In
  \bibinfo{booktitle}{\emph{Proceedings of the 16th Conference of the European
  Chapter of the Association for Computational Linguistics: Main Volume}}.
  \bibinfo{pages}{1075--1088}.
\newblock


\bibitem[\protect\citeauthoryear{Miller, Taori, Raghunathan, Sagawa, Koh,
  Shankar, Liang, Carmon, and Schmidt}{Miller et~al\mbox{.}}{2021}]%
        {Miller2021AccuracyOT}
\bibfield{author}{\bibinfo{person}{John Miller}, \bibinfo{person}{Rohan Taori},
  \bibinfo{person}{Aditi Raghunathan}, \bibinfo{person}{Shiori Sagawa},
  \bibinfo{person}{Pang~Wei Koh}, \bibinfo{person}{Vaishaal Shankar},
  \bibinfo{person}{Percy Liang}, \bibinfo{person}{Yair Carmon}, {and}
  \bibinfo{person}{Ludwig Schmidt}.} \bibinfo{year}{2021}\natexlab{}.
\newblock \showarticletitle{Accuracy on the Line: on the Strong Correlation
  Between Out-of-Distribution and In-Distribution Generalization}.
\newblock  (\bibinfo{year}{2021}).
\newblock


\bibitem[\protect\citeauthoryear{Nguyen, Rosenberg, Song, Gao, Tiwary,
  Majumder, and Deng}{Nguyen et~al\mbox{.}}{2016}]%
        {nguyen2016ms}
\bibfield{author}{\bibinfo{person}{Tri Nguyen}, \bibinfo{person}{Mir
  Rosenberg}, \bibinfo{person}{Xia Song}, \bibinfo{person}{Jianfeng Gao},
  \bibinfo{person}{Saurabh Tiwary}, \bibinfo{person}{Rangan Majumder}, {and}
  \bibinfo{person}{Li Deng}.} \bibinfo{year}{2016}\natexlab{}.
\newblock \showarticletitle{MS MARCO: A human generated machine reading
  comprehension dataset}.
\newblock \bibinfo{journal}{\emph{choice}}  \bibinfo{volume}{2640}
  (\bibinfo{year}{2016}), \bibinfo{pages}{660}.
\newblock


\bibitem[\protect\citeauthoryear{Nogueira and Lin}{Nogueira and Lin}{2019}]%
        {nogueira2019doc2query}
\bibfield{author}{\bibinfo{person}{Rodrigo Nogueira} {and}
  \bibinfo{person}{Jimmy Lin}.} \bibinfo{year}{2019}\natexlab{}.
\newblock \bibinfo{title}{From doc2query to {docTTTTTquery}}.
\newblock
\newblock


\bibitem[\protect\citeauthoryear{Raiber and Kurland}{Raiber and
  Kurland}{2014}]%
        {raiber2014query}
\bibfield{author}{\bibinfo{person}{Fiana Raiber} {and} \bibinfo{person}{Oren
  Kurland}.} \bibinfo{year}{2014}\natexlab{}.
\newblock \showarticletitle{Query-performance prediction: setting the
  expectations straight}. In \bibinfo{booktitle}{\emph{Proceedings of the 37th
  international ACM SIGIR conference on Research \& development in information
  retrieval}}. \bibinfo{pages}{13--22}.
\newblock


\bibitem[\protect\citeauthoryear{Reimers and Gurevych}{Reimers and
  Gurevych}{2019}]%
        {reimers-2019-sentence-bert}
\bibfield{author}{\bibinfo{person}{Nils Reimers} {and} \bibinfo{person}{Iryna
  Gurevych}.} \bibinfo{year}{2019}\natexlab{}.
\newblock \showarticletitle{Sentence-BERT: Sentence Embeddings using Siamese
  BERT-Networks}. In \bibinfo{booktitle}{\emph{Proceedings of the 2019
  Conference on Empirical Methods in Natural Language Processing}}.
  \bibinfo{publisher}{Association for Computational Linguistics}.
\newblock
\urldef\tempurl%
\url{http://arxiv.org/abs/1908.10084}
\showURL{%
\tempurl}


\bibitem[\protect\citeauthoryear{Ren, Qu, Liu, Zhao, Wu, Ding, Wu, Wang, and
  Wen}{Ren et~al\mbox{.}}{2022}]%
        {ren2022thorough}
\bibfield{author}{\bibinfo{person}{Ruiyang Ren}, \bibinfo{person}{Yingqi Qu},
  \bibinfo{person}{Jing Liu}, \bibinfo{person}{Wayne~Xin Zhao},
  \bibinfo{person}{Qifei Wu}, \bibinfo{person}{Yuchen Ding},
  \bibinfo{person}{Hua Wu}, \bibinfo{person}{Haifeng Wang}, {and}
  \bibinfo{person}{Ji-Rong Wen}.} \bibinfo{year}{2022}\natexlab{}.
\newblock \showarticletitle{A thorough examination on zero-shot dense
  retrieval}.
\newblock \bibinfo{journal}{\emph{arXiv preprint arXiv:2204.12755}}
  (\bibinfo{year}{2022}).
\newblock


\bibitem[\protect\citeauthoryear{Rieck, Togninalli, Bock, Moor, Horn, Gumbsch,
  and Borgwardt}{Rieck et~al\mbox{.}}{2019}]%
        {Rieck19a}
\bibfield{author}{\bibinfo{person}{Bastian Rieck}, \bibinfo{person}{Matteo
  Togninalli}, \bibinfo{person}{Christian Bock}, \bibinfo{person}{Michael
  Moor}, \bibinfo{person}{Max Horn}, \bibinfo{person}{Thomas Gumbsch}, {and}
  \bibinfo{person}{Karsten Borgwardt}.} \bibinfo{year}{2019}\natexlab{}.
\newblock \showarticletitle{Neural Persistence: {A} Complexity Measure for Deep
  Neural Networks Using Algebraic Topology}. In
  \bibinfo{booktitle}{\emph{International Conference on Learning
  Representations}}.
\newblock


\bibitem[\protect\citeauthoryear{Saito, Kim, Teterwak, Sclaroff, Darrell, and
  Saenko}{Saito et~al\mbox{.}}{2021}]%
        {9711410}
\bibfield{author}{\bibinfo{person}{K. Saito}, \bibinfo{person}{D. Kim},
  \bibinfo{person}{P. Teterwak}, \bibinfo{person}{S. Sclaroff},
  \bibinfo{person}{T. Darrell}, {and} \bibinfo{person}{K. Saenko}.}
  \bibinfo{year}{2021}\natexlab{}.
\newblock \showarticletitle{Tune it the Right Way: Unsupervised Validation of
  Domain Adaptation via Soft Neighborhood Density}. In
  \bibinfo{booktitle}{\emph{IEEE/CVF International Conference on Computer
  Vision (ICCV)}}. \bibinfo{publisher}{IEEE Computer Society},
  \bibinfo{pages}{9164--9173}.
\newblock


\bibitem[\protect\citeauthoryear{Thakur, Reimers, R{\"u}ckl{\'e}, Srivastava,
  and Gurevych}{Thakur et~al\mbox{.}}{2021}]%
        {thakur2021beir}
\bibfield{author}{\bibinfo{person}{Nandan Thakur}, \bibinfo{person}{Nils
  Reimers}, \bibinfo{person}{Andreas R{\"u}ckl{\'e}}, \bibinfo{person}{Abhishek
  Srivastava}, {and} \bibinfo{person}{Iryna Gurevych}.}
  \bibinfo{year}{2021}\natexlab{}.
\newblock \showarticletitle{BEIR: A Heterogeneous Benchmark for Zero-shot
  Evaluation of Information Retrieval Models}. In
  \bibinfo{booktitle}{\emph{Thirty-fifth Conference on Neural Information
  Processing Systems Datasets and Benchmarks Track (Round 2)}}.
\newblock


\bibitem[\protect\citeauthoryear{Turtle and Croft}{Turtle and Croft}{1997}]%
        {turtle1997uncertainty}
\bibfield{author}{\bibinfo{person}{Howard~R Turtle} {and}
  \bibinfo{person}{W~Bruce Croft}.} \bibinfo{year}{1997}\natexlab{}.
\newblock \showarticletitle{Uncertainty in information retrieval systems}.
\newblock \bibinfo{journal}{\emph{Uncertainty management in information
  systems: from needs to solutions}} (\bibinfo{year}{1997}),
  \bibinfo{pages}{189--224}.
\newblock


\bibitem[\protect\citeauthoryear{Wang and Zhu}{Wang and Zhu}{2009}]%
        {wang2009portfolio}
\bibfield{author}{\bibinfo{person}{Jun Wang} {and} \bibinfo{person}{Jianhan
  Zhu}.} \bibinfo{year}{2009}\natexlab{}.
\newblock \showarticletitle{Portfolio theory of information retrieval}. In
  \bibinfo{booktitle}{\emph{Proceedings of the 32nd international ACM SIGIR
  conference on Research and development in information retrieval}}.
  \bibinfo{pages}{115--122}.
\newblock


\bibitem[\protect\citeauthoryear{Wang, Thakur, Reimers, and Gurevych}{Wang
  et~al\mbox{.}}{2022a}]%
        {wang2022gpl}
\bibfield{author}{\bibinfo{person}{Kexin Wang}, \bibinfo{person}{Nandan
  Thakur}, \bibinfo{person}{Nils Reimers}, {and} \bibinfo{person}{Iryna
  Gurevych}.} \bibinfo{year}{2022}\natexlab{a}.
\newblock \showarticletitle{GPL: Generative Pseudo Labeling for Unsupervised
  Domain Adaptation of Dense Retrieval}. In
  \bibinfo{booktitle}{\emph{Proceedings of the 2022 Conference of the North
  American Chapter of the Association for Computational Linguistics: Human
  Language Technologies}}. \bibinfo{pages}{2345--2360}.
\newblock


\bibitem[\protect\citeauthoryear{Wang, Yang, Huang, Jiao, Yang, Jiang,
  Majumder, and Wei}{Wang et~al\mbox{.}}{2022b}]%
        {wang2022simlm}
\bibfield{author}{\bibinfo{person}{Liang Wang}, \bibinfo{person}{Nan Yang},
  \bibinfo{person}{Xiaolong Huang}, \bibinfo{person}{Binxing Jiao},
  \bibinfo{person}{Linjun Yang}, \bibinfo{person}{Daxin Jiang},
  \bibinfo{person}{Rangan Majumder}, {and} \bibinfo{person}{Furu Wei}.}
  \bibinfo{year}{2022}\natexlab{b}.
\newblock \showarticletitle{SimLM: Pre-training with Representation Bottleneck
  for Dense Passage Retrieval}.
\newblock \bibinfo{journal}{\emph{arXiv preprint arXiv:2207.02578}}
  (\bibinfo{year}{2022}).
\newblock


\bibitem[\protect\citeauthoryear{Xiong, Xiong, Li, Tang, Liu, Bennett, Ahmed,
  and Overwijk}{Xiong et~al\mbox{.}}{[n.d.]}]%
        {xiongapproximate}
\bibfield{author}{\bibinfo{person}{Lee Xiong}, \bibinfo{person}{Chenyan Xiong},
  \bibinfo{person}{Ye Li}, \bibinfo{person}{Kwok-Fung Tang},
  \bibinfo{person}{Jialin Liu}, \bibinfo{person}{Paul~N Bennett},
  \bibinfo{person}{Junaid Ahmed}, {and} \bibinfo{person}{Arnold Overwijk}.}
  \bibinfo{year}{[n.d.]}\natexlab{}.
\newblock \showarticletitle{Approximate Nearest Neighbor Negative Contrastive
  Learning for Dense Text Retrieval}. In
  \bibinfo{booktitle}{\emph{International Conference on Learning
  Representations}}.
\newblock


\bibitem[\protect\citeauthoryear{Yates, Nogueira, and Lin}{Yates
  et~al\mbox{.}}{2021}]%
        {yates2021pretrained}
\bibfield{author}{\bibinfo{person}{Andrew Yates}, \bibinfo{person}{Rodrigo
  Nogueira}, {and} \bibinfo{person}{Jimmy Lin}.}
  \bibinfo{year}{2021}\natexlab{}.
\newblock \showarticletitle{Pretrained transformers for text ranking: BERT and
  beyond}. In \bibinfo{booktitle}{\emph{Proceedings of the 14th ACM
  International Conference on web search and data mining}}.
  \bibinfo{pages}{1154--1156}.
\newblock


\bibitem[\protect\citeauthoryear{Yilmaz, Aslam, and Robertson}{Yilmaz
  et~al\mbox{.}}{2008}]%
        {yilmaz2008a-new-rank}
\bibfield{author}{\bibinfo{person}{Emine Yilmaz}, \bibinfo{person}{Javed~A.
  Aslam}, {and} \bibinfo{person}{Stephen Robertson}.}
  \bibinfo{year}{2008}\natexlab{}.
\newblock \showarticletitle{A New Rank Correlation Coefficient for Information
  Retrieval}. In \bibinfo{booktitle}{\emph{Proceedings of the 31st Annual
  International ACM SIGIR Conference on Research and Development in Information
  Retrieval}} (Singapore, Singapore) \emph{(\bibinfo{series}{SIGIR '08})}.
  \bibinfo{publisher}{Association for Computing Machinery},
  \bibinfo{address}{New York, NY, USA}, \bibinfo{pages}{587--594}.
\newblock
\showISBNx{9781605581644}
\urldef\tempurl%
\url{https://doi.org/10.1145/1390334.1390435}
\showDOI{\tempurl}


\bibitem[\protect\citeauthoryear{You, Wang, Long, and Jordan}{You
  et~al\mbox{.}}{2019}]%
        {pmlr-v97-you19a}
\bibfield{author}{\bibinfo{person}{Kaichao You}, \bibinfo{person}{Ximei Wang},
  \bibinfo{person}{Mingsheng Long}, {and} \bibinfo{person}{Michael Jordan}.}
  \bibinfo{year}{2019}\natexlab{}.
\newblock \showarticletitle{Towards Accurate Model Selection in Deep
  Unsupervised Domain Adaptation}. In \bibinfo{booktitle}{\emph{Proceedings of
  the 36th International Conference on Machine Learning}}
  \emph{(\bibinfo{series}{Proceedings of Machine Learning Research})},
  \bibfield{editor}{\bibinfo{person}{Kamalika Chaudhuri} {and}
  \bibinfo{person}{Ruslan Salakhutdinov}} (Eds.), Vol.~\bibinfo{volume}{97}.
  \bibinfo{publisher}{PMLR}, \bibinfo{pages}{7124--7133}.
\newblock
\urldef\tempurl%
\url{https://proceedings.mlr.press/v97/you19a.html}
\showURL{%
\tempurl}


\bibitem[\protect\citeauthoryear{Zhao, Liu, Ren, and Wen}{Zhao
  et~al\mbox{.}}{2022}]%
        {zhao2022dense}
\bibfield{author}{\bibinfo{person}{Wayne~Xin Zhao}, \bibinfo{person}{Jing Liu},
  \bibinfo{person}{Ruiyang Ren}, {and} \bibinfo{person}{Ji-Rong Wen}.}
  \bibinfo{year}{2022}\natexlab{}.
\newblock \showarticletitle{Dense text retrieval based on pretrained language
  models: A survey}.
\newblock \bibinfo{journal}{\emph{arXiv preprint arXiv:2211.14876}}
  (\bibinfo{year}{2022}).
\newblock


\bibitem[\protect\citeauthoryear{Zhou and Croft}{Zhou and Croft}{2007}]%
        {zhou2007query}
\bibfield{author}{\bibinfo{person}{Yun Zhou} {and} \bibinfo{person}{W~Bruce
  Croft}.} \bibinfo{year}{2007}\natexlab{}.
\newblock \showarticletitle{Query performance prediction in web search
  environments}. In \bibinfo{booktitle}{\emph{Proceedings of the 30th annual
  international ACM SIGIR conference on Research and development in information
  retrieval}}. \bibinfo{pages}{543--550}.
\newblock


\bibitem[\protect\citeauthoryear{Zhu, Wang, Cox, and Taylor}{Zhu
  et~al\mbox{.}}{2009}]%
        {zhu2009risky}
\bibfield{author}{\bibinfo{person}{Jianhan Zhu}, \bibinfo{person}{Jun Wang},
  \bibinfo{person}{Ingemar~J Cox}, {and} \bibinfo{person}{Michael~J Taylor}.}
  \bibinfo{year}{2009}\natexlab{}.
\newblock \showarticletitle{Risky business: modeling and exploiting uncertainty
  in information retrieval}. In \bibinfo{booktitle}{\emph{Proceedings of the
  32nd international ACM SIGIR conference on Research and development in
  information retrieval}}. \bibinfo{pages}{99--106}.
\newblock


\bibitem[\protect\citeauthoryear{Zhuang and Zuccon}{Zhuang and Zuccon}{2021}]%
        {zhuang2021dealing}
\bibfield{author}{\bibinfo{person}{Shengyao Zhuang} {and}
  \bibinfo{person}{Guido Zuccon}.} \bibinfo{year}{2021}\natexlab{}.
\newblock \showarticletitle{Dealing with Typos for {BERT}-based Passage
  Retrieval and Ranking}. In \bibinfo{booktitle}{\emph{Proceedings of the 2021
  Conference on Empirical Methods in Natural Language Processing}}.
  \bibinfo{publisher}{Association for Computational Linguistics},
  \bibinfo{address}{Online and Punta Cana, Dominican Republic},
  \bibinfo{pages}{2836--2842}.
\newblock


\bibitem[\protect\citeauthoryear{Zhuang and Zuccon}{Zhuang and Zuccon}{2022}]%
        {zhuang2022char}
\bibfield{author}{\bibinfo{person}{Shengyao Zhuang} {and}
  \bibinfo{person}{Guido Zuccon}.} \bibinfo{year}{2022}\natexlab{}.
\newblock \showarticletitle{CharacterBERT and Self-Teaching for Improving the
  Robustness of Dense Retrievers on Queries with Typos}
  \emph{(\bibinfo{series}{SIGIR '22})}. \bibinfo{publisher}{Association for
  Computing Machinery}, \bibinfo{address}{New York, NY, USA},
  \bibinfo{pages}{1444--1454}.
\newblock


\bibitem[\protect\citeauthoryear{Zuccon, Azzopardi, and Van~Rijsbergen}{Zuccon
  et~al\mbox{.}}{2009}]%
        {zuccon2009quantum}
\bibfield{author}{\bibinfo{person}{Guido Zuccon}, \bibinfo{person}{Leif~A
  Azzopardi}, {and} \bibinfo{person}{Keith Van~Rijsbergen}.}
  \bibinfo{year}{2009}\natexlab{}.
\newblock \showarticletitle{The quantum probability ranking principle for
  information retrieval}. In \bibinfo{booktitle}{\emph{Advances in Information
  Retrieval Theory: Second International Conference on the Theory of
  Information Retrieval, ICTIR 2009 Cambridge, UK, September 10-12, 2009
  Proceedings 2}}. Springer, \bibinfo{pages}{232--240}.
\newblock


\end{thebibliography}

\end{document}